\documentclass[12pt]{iopart}
\usepackage[slovene,english]{babel} 
\usepackage{graphicx,bbm,amssymb,amsopn}
\usepackage{dsfont}

\newcommand{\be}{\begin{equation}}
\newcommand{\ee}{\end{equation}}

\newcommand{\ave}[1]{{\langle #1\rangle}}
\newcommand{\ii}{ {\rm i} }
\newcommand{\dd}{ {\rm d} }

\newcommand{\mm}[1]{{\mathbf{#1}}}

\def\tr{{{\rm tr}}}

\begin{document}

\title{Lindblad master equation approach to superconductivity in open quantum systems}
 
\author{Daniel S. Kosov${}^1$, Toma\v z Prosen${}^2$, and Bojan \v{Z}unkovi\v{c}${}^{2}$}

\address{
${}^1$ Department of Physics, 
Universit\'e Libre de Bruxelles, Campus Plaine, CP 231, Blvd du Triomphe, B-1050 Brussels, Belgium\\
${}^2$Department of Physics, Faculty of Mathematics and Physics, University of Ljubljana, Jadranska 19, SI-1000 Ljubljana, Slovenia}

\date{\today}

\begin{abstract}
We consider an open quantum Fermi-system which consists of a  single degenerate level with pairing interactions embedded into 
a superconducting bath.  The time evolution of the reduced density matrix for the system is given
by Linblad master equation, where the dissipators describe exchange of Bogoliubov quasiparticles with the bath.
We obtain  fixed points of the time evolution equation for the covariance matrix and study their stability by analyzing  full dynamics of 
the order parameter.

\end{abstract}
~~~~~~~~~~~~PACS numbers: 74.20.Fg, 03.65.Yz

\maketitle

The Bardeen, Cooper, and Schrieffer (BCS) theory of superconductivity \cite{bcs57}  is based on the simple Hamiltonian but it captures the essential physics not only  for superconductivity of electrons in metals,  but also the superconductivity of atomic nuclei, nuclear matter,   neutron stars \cite{RevModPhys.75.607}, and cold atomic Fermi-gases \cite{Gurarie20072}. 
Here, based on the Lindblad master equation, we propose the extension of the BCS theory to the open quantum system. We determine the order parameter -- the density of Cooper pairs, and optionally the order parameters of a superconducting reservoir, self consistently. A simple model of a single spinful fermionic level embedded into a fermionic reservoir is proposed. 
For a self consistent treatment of the reservoir we recover the standard mean-field superconducting phase transition in the grand-canonical (equilibrium) state, whereas for a fixed state of the reservoir, we find that the
state of the system follows the state of the reservoir, being either superconducting or normal.

We consider a Lindblad master equation for the time-dependent reduced density matrix $\rho(t)$ \cite{lindblad76}
\begin{eqnarray}
\label{eq:lindblad}
\frac{{\rm d}\rho}{{\rm dt}}=-{\rm i}[H,\rho]+\sum_{\nu=1}^{M}\left(2L_\nu\rho L_\nu^\dag -\{L^\dag_\nu L_\nu,\rho\}\right),
\end{eqnarray}
where Hamiltonian $H$  is given as
\begin{eqnarray}
\label{eq:ExactH}
H=\epsilon(a_{\uparrow}^\dag a_{\uparrow}+a_{\downarrow}^\dag a_{\downarrow})+ U a_{\uparrow}^\dag a_{\downarrow}^\dag a_{\downarrow} a_{\uparrow}.
\end{eqnarray}
The Hamiltonian consists of a single degenerate level with BCS type pairing interaction;  $a^\dagger_\sigma,a_\sigma$ are standard fermionic creation/annihilation operators ($\sigma=\uparrow,\downarrow$), and $L_\mu$ are some Lindblad operators which will be specified later.
Introducing (time-dependent) averages $\ave{A} = \tr A \rho(t)$, and using Wick theorem we approximate (\ref{eq:ExactH}) by the 
mean-field Bogoliubov-de Gennes Hamiltonian
\begin{eqnarray}
\label{eq:H_def}
H=\epsilon(a_{\uparrow}^\dag a_{\uparrow}+a_{\downarrow}^\dag a_{\downarrow})+\Delta({\rm e}^{{\rm i} \chi}a_{\downarrow} a_{\uparrow}+{\rm e}^{-{\rm i} \chi}a_{\uparrow}^\dag a_{\downarrow}^\dag),
\end{eqnarray}
where 
\begin{equation}
\Delta {\rm e}^{\ii \chi} = U \ave{a^\dag_\uparrow a^\dag_\downarrow}
\label{eq:Delta}
\end{equation}
is the complex order parameter.

The Hamiltonian (\ref{eq:H_def}) can be diagonalized by the canonical Bogoliubov transformation
\begin{eqnarray}
\alpha_1=-{\rm e}^{{\rm i}\chi}\cos(\phi)a_{\uparrow}+\sin(\phi)a_{\downarrow}^\dag, \\
%\quad &\alpha_1^\dag=-{\rm e}^{-{\rm i}\chi}\cos(\phi)a_{\uparrow}^\dag+\sin(\phi)a_{\downarrow},\\ \nonumber
%\alpha_2={\rm e}^{{\rm i}\chi}\cos(\phi)a_{\downarrow}+\sin(\phi)a_{\uparrow}^\dag,\quad &
\alpha_2^\dag={\rm e}^{-{\rm i}\chi}\cos(\phi)a_{\downarrow}^\dag+\sin(\phi)a_{\uparrow},
\end{eqnarray}
where $\tan(2 \phi)=-\frac{\Delta}{\epsilon}$, and $\alpha_i$ are Bogoliubov quasiparticles, which satisfy standard anticommutation relations.
%$\{\alpha^\dagger_j,\alpha_k\}=\delta_{j,k}$, $\{\alpha_j,\alpha_k\}=0$, $j,k=1,2$ 
The diagonal form of the Hamiltonian is
\begin{eqnarray}
H= \sqrt{\epsilon^2+\Delta^2} (\alpha_1^\dag\alpha_1+\alpha_2^\dag\alpha_2).
\end{eqnarray}
We assume that our system is embedded into a superconducting bath described by a macroscopic gas of Bogoliubov quasiparticles, which can be exchanged with the system. This process is quite generally described by a combination of the following $M=4$ Lindblad operators
\begin{eqnarray}
L_1=\sqrt{\Gamma_{1}}\left(-{\rm e}^{{\rm i}\eta}\cos(\theta)a_{\uparrow}+\sin(\theta)a_{\downarrow}^\dag \right),\\ \nonumber
L_2=\sqrt{\Gamma_{2}}\left(-{\rm e}^{-{\rm i}\eta}\cos(\theta)a_{\uparrow}^\dag+\sin(\theta)a_{\downarrow} \right),\\ \nonumber
L_3=\sqrt{\Gamma_{1}}\left({\rm e}^{{\rm i}\eta}\cos(\theta)a_{\downarrow}+\sin(\theta)a_{\uparrow}^\dag \right),\\ \nonumber
L_4=\sqrt{\Gamma_{2}}\left({\rm e}^{-{\rm i}\eta}\cos(\theta)a_{\downarrow}^\dag+\sin(\theta)a_{\uparrow} \right).
\end{eqnarray}
Superconducting properties of the bath are not necessarily the same as of the embedded system, so the angles $\theta$, and $\eta$, may be different
from $\phi$, and $\chi$, respectively. The coupling constants $\Gamma_{1,2}$ are connected to the Fermi-Dirac distribution of the normal modes of the Bogoliubov-de Gennes Hamiltonian
(\ref{eq:H_def}):
\begin{eqnarray}
\Gamma_1=\gamma(1-f),\quad\Gamma_2=\gamma f, \quad f =\frac{1}{1+{\rm e}^{\beta \sqrt{\epsilon^2+\Delta^2}}},
\label{eq:Gamma}
\end{eqnarray}
where $\gamma$ is a parameter controlling the strength of the system-bath coupling.

It is convenient to rewrite the Lindblad equation in terms of Hermitian Majorana fermions 
\begin{eqnarray}
\label{eq:transf}
w_1=a_\uparrow+a_\uparrow^\dag, \quad w_2={\rm i}(a_\uparrow-a_\uparrow^\dag), \quad w_3= a_\downarrow+a_\downarrow^\dag,
\quad w_4={\rm i}(a_\downarrow-a_\downarrow^\dag),
\end{eqnarray}
satisfying $\{ w_j, w_k\} = 2 \delta_{j,k}$, $j,k=1,\ldots,4$.
Since our master equation is quadratic in terms of $w_j$, we obtain a closed set of equations for the covariance matrix 
$\ave{w_j w_k} = \delta_{j,k} - \ii Z_{j,k}(t)$:
\begin{eqnarray}
\label{eq:diff_eq}
\frac{{\rm d}{\bf Z}}{{\rm d}t} = -{\bf X}^{\rm T}{\bf Z} - {\bf ZX} + {\bf Y}.
\end{eqnarray}
Here  $\mm{Z}(t)$ is a real, anti-symmetric $4\times 4$ matrix,
$\mm{X}$ and $\mm{Y}$ are real matrices
\begin{eqnarray}
&&\mm{X}=
\pmatrix{
 \gamma  & -\epsilon  & -\Delta \sin \chi & \Delta  \cos\chi \cr
 \epsilon  & \gamma  & \Delta \cos \chi & \Delta  \sin\chi \cr
 \Delta \sin\chi & -\Delta \cos\chi & \gamma  & -\epsilon  \cr
 -\Delta \cos\chi & -\Delta \sin\chi & \epsilon  & \gamma 
}, \label{eq:X} \\
&&\!\!\!\!\!\!\!\!\!\!\!\!\!\!\!\!\!\!\!\!\!\!\!\!\!\!\!\!\!\mm{Y} = 
2\gamma (1-2f)
\pmatrix{
 0 & \cos 2 \theta & -  \sin\eta\sin2\theta  &  \cos\eta\sin2\theta \cr
 - \cos2\theta & 0 & \cos\eta\sin2\theta &  \sin\eta\sin2\theta \cr
  \sin\eta\sin2\theta & -\cos\eta\sin2\theta & 0 &  \cos2\theta \cr
 -\cos\eta\sin2\theta & -\sin\eta\sin 2\theta & -\cos2\theta & 0
}.
\end{eqnarray}
Note that eq. (\ref{eq:diff_eq}) is non-linear, as $\Delta$ and $\chi$ depend again on covariances through the relation (\ref{eq:Delta}).
If in addition, we determine the Lindblad operators self-consistently, which means that
the bath has the same properties as the embedded system, then we should also set $\tan2\theta = -\frac{\Delta}{\epsilon}$, ($\theta\equiv\phi$), and
$\eta \equiv \chi$.

Fixed points of the flow (\ref{eq:diff_eq}), which are solutions of the continuous Lyapunov equation
\begin{equation}
\mm{X}^T \mm{Z} + \mm{Z}\mm{X} = \mm{Y},
\label{eq:Lyap}
\end{equation}
determine the stationary states of the system. These stationary states  may not be unique because of the non-linearity.
However, one can show that a stable fixed point is unique. It follows from the fact that all linear relaxation rates, i.e. eigenvalues of the matrix $\mm{X}$ (\ref{eq:X}), $x_{1,2,3,4} = \gamma \pm \ii \sqrt{\Delta^2+\epsilon^2}$, have strictly positive real parts for $\gamma > 0$.
One can further show that all solutions of Eq. (\ref{eq:Lyap}) are of the form
\begin{equation}
\mm{Z} = \pmatrix{
 0 &  z_1 &  z_2 &  z_3 \cr
 - z_1 & 0 &  z_3 & - z_2 \cr
 - z_2 & - z_3 & 0 &  z_1 \cr
 - z_3 &  z_2 & - z_1 & 0
}
\label{eq:Z}
\end{equation}
which is specified by only three real variables $z_1,z_2,z_3$. 
The Lyapunov equation should be solved self-consistently (\ref{eq:Delta}), i.e.
\begin{equation}
\label{eq:consistent_condition}
z_2 = -\frac{2\Delta}{U} \sin \chi,
\quad
z_3 = \frac{2\Delta}{U} \cos \chi.
\label{eq:SelfC}
\end{equation}
Let us now consider two possible cases of the bath, i.e. two possible choices of angles $\theta,\eta$:

\begin{figure}
          \centering	
	\includegraphics[width=0.65\columnwidth]{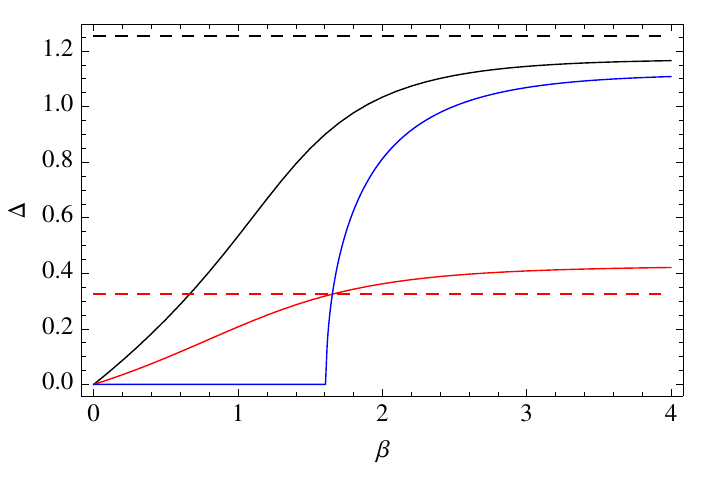}
	\caption{The phase diagram of the magnitude of the order parameter $\Delta$ versus the inverse temperature. $\Delta > 0$ signals the superconducting phase. Blue curve corresponds to self-consistently determined bath, whereas black and red curves correspond to fixed superconducting baths, with
	$\Delta_{\rm bath} = -\epsilon \tan 2\theta$ (indicated by horizontal dashed lines), with $\theta =-\pi/20$ and $\theta= -\pi/7$, respectively. Other parameters are $\epsilon=1,\gamma=1,U= - 3$.  }
	\label{fig:phase}
\end{figure}

\noindent
(i) {\em Fixed bath.}
If the bath is considered fixed, we can set $\eta:=0$ without loss of generality, so the equation \eref{eq:SelfC} together with the Lyapunov equation (\ref{eq:Lyap}) results in the conditions
\begin{eqnarray}
\!\!\!\!\!\!\!\!\!\!\!\!\!\!\!\!\!\!&&(\gamma\sin\chi - \epsilon\cos\chi) \sin2\theta  = \Delta  \cos2\theta, \\ \nonumber
\!\!\!\!\!\!\!\!\!\!\!\!\!\!\!\!\!\!&&
U(1-2f) \left(\left(\left(\gamma ^2+\Delta ^2\right) \cos\chi + \gamma   \epsilon\sin\chi\right)\sin2\theta  -  \Delta\epsilon\cos2\theta    \right) = 2 \Delta  \left(\gamma ^2+\Delta^2+\epsilon^2\right) .
   \label{eq:fixedbath}
\end{eqnarray} 
For normal, non-superconducting bath, $\sin2 \theta=0$, and  there exists only a trivial solution $\Delta=0$ for the system.
For a superconducting bath, $\sin2\theta\neq 0$, the system is always superconducting as well, for all inverse temperatures $\beta$, as the only solution  (\ref{eq:fixedbath}) has $\Delta \neq 0$. In other words, for a superconducting bath the embedded system cannot be in the normal state no matter how small the parameter $\gamma$ is.

\noindent 
(ii) {\em Self-consistent bath.}
The bath can  be considered self-consistent in the following way. We associate the Lindblad operators with the normal mode  creation and annihilation operators such that $\theta=\phi$, $\eta=\chi$. 
Inserting these assumptions in the Lyapunov equation \eref{eq:Lyap} we get
\begin{eqnarray}
\label{eq:z_solution}
z_1=\frac{\epsilon  \tanh \left(\frac{1}{2} \beta  \sqrt{\Delta ^2+\epsilon
   ^2}\right)}{ \sqrt{\Delta ^2+\epsilon ^2}},\\ \nonumber
z_2=\frac{\Delta  \sin (\chi ) \tanh \left(\frac{1}{2} \beta  \sqrt{\Delta
   ^2+\epsilon ^2}\right)}{\sqrt{\Delta ^2+\epsilon ^2}},\\ \nonumber
z_3=-\frac{\Delta  \cos (\chi ) \tanh \left(\frac{1}{2} \beta  \sqrt{\Delta
   ^2+\epsilon ^2}\right)}{ \sqrt{\Delta ^2+\epsilon ^2}}.
\end{eqnarray}
The consistency conditions  \eref{eq:SelfC} now result in
\begin{eqnarray}
\label{eq:condition2}
 \left(\frac{U \tanh \left(\frac{1}{2} \beta 
   \sqrt{\Delta ^2+\epsilon ^2}\right)}{2\sqrt{\Delta ^2 +\epsilon ^2}}+1\right) \Delta  = 0
\end{eqnarray}
We always have the trivial solution $\Delta=0$, however for temperatures smaller than $1/\beta_{\rm c}$ where
\begin{eqnarray}
\label{eq:betac}
\beta_{\rm c}=-\frac{2}{\epsilon} {\rm artanh} \left(\frac{ 2\epsilon }{U}\right) 
\end{eqnarray}
we also find superconducting solution with $\Delta \neq 0$. Obviously, critical temperature only exists for $-U > 2\epsilon$.  Eqs. (\ref{eq:condition2}, \ref{eq:betac}) are exactly the result
that we get in standard grand canonical ensemble. The bath-self-consistency conditions completely eliminate dependence of the properties of the stationary state on the system-bath coupling $\gamma$.

In figure~\ref{fig:phase} we plot a phase diagram $\Delta(\beta)$, for both cases (i,ii).  Note that only if 
$\Delta_{\rm bath}:=-\epsilon\tan2\theta$ is smaller than $ \sqrt{\frac{1}{4}U^2-\epsilon^2}$ then we may have $\Delta=\Delta_{\rm bath}$ for some temperature 
$1/\beta$.

\begin{figure}
          \centering	
	\includegraphics[width=0.65\columnwidth]{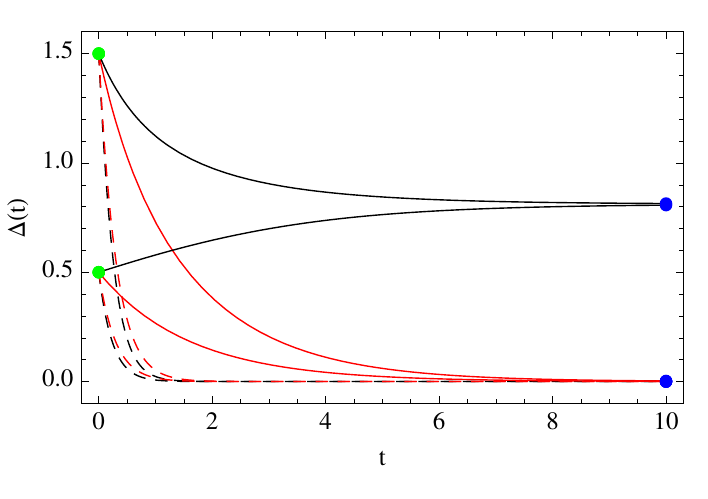}
	\caption{Time-dependent order parameter $\Delta(t)$ as a function of time, for $\beta < \beta_{\rm c}$ (red curves) and $\beta > \beta_{\rm c}$ (black curves), and for an attractive potential $U < 0$ (full curves) and a repulsive potential $U > 0$ (dashed curves). We plot some typical trajectories starting from green points and ending in one of the two stable fixed points (blue).  Numerical values of parameters are $\epsilon=1,\gamma=1,|U|=3$.}
		\label{fig:dynamic}
\end{figure}

So far we have  investigated fixed points of the non-linear flow (\ref{eq:diff_eq}). Let us now address the question of their stability by investigating the full dynamics. We shall only focus on the case (ii) of self-consitently determined baths, which possesses non-unique stationary solution below the critical temperature. The dynamical equations for $z_j(t), j=1,2,3$ follow directly from equation (\ref{eq:diff_eq}) with the ansatz (\ref{eq:Z}). 
By means of self-consistency equations (\ref{eq:SelfC}) we replace $z_2(t),z_3(t)$ by $\Delta(t),\chi(t)$, resulting in a system of three non-linear differential equations
\begin{eqnarray}
\frac{\dd\Delta}{\dd t} 
=-2\gamma \Delta\left(\frac{U\tanh\left( \frac{\beta}{2}\sqrt{\epsilon^2+\Delta^2}\right)}{2\sqrt{\epsilon^2+\Delta^2}} +1 \right), \label{eq:Dt} \\
\frac{\dd{z_1}}{\dd t}= 2\gamma\left(\epsilon \frac{\tanh\left( \frac{\beta}{2}\sqrt{\epsilon^2+\Delta^2}\right)}{\sqrt{\epsilon^2+\Delta^2}} -z_1\right),\label{eq:z1t}\\  
\frac{\dd\chi}{\dd t} =U z_1 + 2\epsilon. \label{eq:chit}
\end{eqnarray}
Note that the first equation (\ref{eq:Dt}) is independent from the other two, and yields a closed first order differential equation for the magnitude of the order parameter $\Delta$. Writing it as $ \dd \Delta/\dd t = G(\Delta)$ we can easily study the stability of two possible 
fixed points  $\Delta_1$, and $\Delta_2$ ($G(\Delta_j)=0$, $j=1,2$, $\Delta_1=0$, and $\Delta_2 \neq 0$ if $\beta > \beta_{\rm c}$). 
For the trivial fixed point we find
\begin{eqnarray}
G'(0)=-2\gamma\left( \frac{U}{2\epsilon} \tanh \left(\frac{\beta  \epsilon }{2}\right) + 1\right),
\end{eqnarray}
which means that the non-superconducting state is stable $G'(0) < 0$ if $\beta < \beta_{\rm c}$ and unstable $G'(0) > 0$ if $\beta > \beta_{\rm c}$.
As for the second, non-trivial fixed point $\Delta_2$, one can easily show that it is always stable $G'(\Delta_2) < 0$ if it exists, i.e. if $\beta > \beta_{\rm c}$.
The flow of the order parameter $\Delta(t)$ in different cases is illustrated in Fig.~\ref{fig:dynamic}. We note that the other two phase-space variables $z_1(t),\chi(t)$ are completely {\em enslaved} by the order parameter $\Delta(t)$,
so they again converge either to their trivial (non-superconducting) or non-trivial (superconducting) fixed point values.
   
In conclusion, based on the Lindblad master equation we proposed the extension of the BCS theory to open Fermi systems exchanging Bogouliubov quasiparticles with the bath. We derived the equations of motion for the covariance matrix and found the fixed points of the flow equations. If the bath is considered self-consistently with the system, the results becomes equivalent to the  grand canonical ensemble and all dependences on the Lindblad dissipators are eliminated. 
If the superconductivity of the bath is fixed, the system  remains in the superconducting state for all values of temperature  no matter how small the system-bath coupling is. 
We performed the stability analysis of the fixed points and found that below the critical temperature the fixed point which corresponds to the normal phase is not stable, whereas the superconducting solution is stable. 
Note that our results are closely related to an exact treatment of open BCS model in quasi-spin formulation \cite{buffet}. However, in Ref. \cite{buffet} the (quasi-)particles cannot be exchanged with the system, so the study 
of thermodynamic limit is more subtle. We thus believe that our single-level formulation provides a minimal model of open BCS quantum dynamics and should serve as the first step in approaching the 
non-equilibrium open BCS models with several different temperature/particle reservoirs.

 This work has been supported by the Francqui Foundation, Programme d'Actions de Recherche Concert\'ee de la Communaut\'e francaise (Belgium) under project ``Theoretical and experimental approaches to surface reactions",
 and the grants P1-0044 and J1-2208 of Slovenian Research Agency (ARRS).

\section*{Referneces}

\end{document}